\documentclass[12pt]{article}
\usepackage{latexsym}
\usepackage{amsmath}
\usepackage{bm}
\usepackage{graphicx,color}
\usepackage{amssymb}
\usepackage{cite}
\usepackage{ulem}

\newcommand{\al}{\alpha}
\newcommand{\be}{\beta}
\newcommand{\ga}{\gamma}
\newcommand{\de}{\delta}
\renewcommand{\theequation}{\thesection.\arabic{equation}}
\newcommand{\ee}{{\rm e}}

\makeatletter
\newcommand\xleftrightarrow[2][]{%
  \ext@arrow 9999{\longleftrightarrowfill@}{#1}{#2}}
\newcommand\longleftrightarrowfill@{%
  \arrowfill@\leftarrow\relbar\rightarrow}
\makeatother
\makeatletter
\newcounter{subeqncnt}
\def\thesubeqncnt{\alph{subeqncnt}}
\def\subequations{\begingroup%
\stepcounter{equation}\edef\@tempa{\theequation}%
\let\c@equation\c@subeqncnt\c@subeqncnt\z@
\edef\theequation{\@tempa\noexpand\thesubeqncnt}}

\makeatother
\title{The Construction of the mKdV Cyclic Symmetric
$N$-soliton Solution  by the B\"{a}cklund Transformation}
\author{Masahito Hayashi\thanks{masahito.hayashi@oit.ac.jp}\\
Osaka Institute of Technology, Osaka 535-8585, Japan\\
Kazuyasu Shigemoto\thanks{shigemot@tezukayama-u.ac.jp} \\
Tezukayama University, Nara 631-8501, Japan\\
Takuya Tsukioka\thanks{tsukioka@bukkyo-u.ac.jp}\\
Bukkyo University, Kyoto 603-8301, Japan\\
}
\date{\empty}

\begin{document}

\maketitle
\abstract{
We study group theoretical structures of the mKdV equation.
The Schwarzian type mKdV equation has the global 
M\"{o}bius group symmetry. 
The Miura transformation makes a connection between 
the mKdV equation and the KdV equation. 
We find the special local M\"{o}bius transformation 
on the mKdV one-soliton solution 
which can be regarded as the commutative KdV B\"{a}cklund transformation 
can generate the mKdV cyclic symmetric $N$-soliton solution. 
In this algebraic construction to obtain multi-soliton solutions, 
we could observe the addition formula. 
}

%%%%%%%%%%%%%%%%%%%%%%%%%%%%%%%%%%%%%%%%%
\section{Introduction} 

\setcounter{equation}{0}

The discovery of the soliton~\cite{Gardner,Lax,Zakhrov} has given 
the breakthrough to exactly solve non-linear equations. 
There have been many interesting 
developments to understand soliton systems  
such as the AKNS formulation~\cite{Ablowitz,Sasaki}, 
the B\"{a}cklund
transformation~\cite{Wahlquist,Wadati1,Wadati2,Hirota3}, 
the Hirota equation~\cite{Hirota1,Hirota2,Hirota4}, 
the Sato theory~\cite{Sato},
the vertex construction of the soliton solution~\cite{Lepowsky,Date}, 
and the Schwarzian type mKdV/KdV equation~\cite{Weiss}. 

In this paper, we focus on algebraic soliton systems.  
The algebraic soliton system which we call here is a subclass of soliton
systems in which  soliton equations allow to
construct the cyclic symmetric $N$-soliton solution 
by applying algebraic addition
formula to one-soliton solution without solving differential equations directly. 
The algebraic soliton system can be regarded as a non-linear 
generalization of the linear system. 
In order to construct solutions of linear differential equations, 
a linear superposition 
plays a key role. 
The algebraic soliton equation which is the special type of 
non-linear differential equation allows the algebraic non-linear 
superposition i.e.\ ``B\"{a}cklund transformation''.  

In the AKNS formulation, the mKdV equation comes from the integrability of 
$2\times2$ matrix and M\"obius group GL(2,$\mathbb R$) naturally
appears~\cite{Ablowitz}. 
However,  the group structure behind the 
mKdV equation and its solutions has not yet been well studied. 
In this paper, 
we would like to reveal that the algebraic addition formula in the algebraic 
soliton system is nothing but the local Abelian sub-``gauge''
transformation of the local non-Abelian 
M\"{o}bius (GL(2,$\mathbb{R}$)) ``gauge'' transformation.
More precisely we would like to answer 
the following questions; 
From what kind of special one-soliton solution, 
can we algebraically construct the cyclic symmetric
$N$-soliton solution? What kind of addition structure of 
the M\"{o}bius group appears in the algebraic cyclic symmetric
 $N$-soliton solution?

%%%%%%%%%%%%%%%%%%%%%%%%%%%%%%%%%%%%%%%%%%%%
\section{Various types of mKdV equations 
and algebraic construction of solutions} 

\setcounter{equation}{0}

%%%%%%%%%%%%%%%%%%%%%%%%%%%
\subsection{mKdV equations and global M\"{o}bius group symmetries }

The mKdV equation with the variable $v=w_x$ is given by
\begin{align}
  v_t-v_{xxx}+6 v^2 v_x&=0,
\label{2e1}\\
  w_t-w_{xxx}+2 {w^3_x}&=0,
\label{2e2} 
\end{align}
where we set integration constants to be zero. 
In order to see symmetries of the soliton equations, 
we first rewrite them to the Schwarzian type mKdV equation.  
Introducing new 
variable $\varphi$ through $\varphi_x=\ee^{2w}$~\cite{Weiss} and 
manipulating the following expressions, 
\begin{align*}
   v_t-v_{xxx}+6 v^2 v_x&=\frac{1}{2} \partial_x 
   \Big( \frac{1}{\varphi_x} (\varphi_t-\varphi_x S(\varphi, x))_x\Big)=0, 
\\
   v&=w_x=\frac{1}{2} \frac{\varphi_{xx}} {\varphi_x}
= \frac{1}{2}  (\log{\varphi_x})_x, 
\end{align*}
with the Schwarzian derivative $S(\varphi,x)$ defined as 
\begin{equation}
 S(\varphi,x)= \frac{\varphi_{xxx}}{\varphi_x}
   -\frac{3}{2} \frac{\varphi^2_{xx}} {\varphi^2_x}=-2\sqrt{\varphi_x} 
   \partial^2_x \Big(\frac{1}{\sqrt{\varphi_x} }\Big), 
\label{2e5}
\end{equation}
we arrive at the Schwarzian type mKdV equation 
\begin{equation}
   \frac{\varphi_t}{\varphi_x}=S(\varphi, x), 
\label{2e6}
\end{equation}
where we impose an integration constant to be zero.

We also prepare the Hirota type mKdV equation~\cite{Hirota2,Hirota4}. 
If we put 
\begin{equation}
\tanh{\frac{w}{2}}=\frac{g}{f} \ \Longleftrightarrow \  
 \ee^{w}=\frac{f+g}{f-g}, 
\label{2e7}  
\end{equation}
the standard mKdV equation (\ref{2e2}) becomes 
the following Hirota form 
\begin{equation}
  \frac{(-D_t+D^3_x)f \cdot g}{D_x f \cdot g}
   =3\cdot\frac{D^2_x (f \cdot f-g \cdot g)}{(f^2-g^2)},
\label{2e8}
\end{equation}
where $D_x$ and $D_t$ are Hirota derivatives. 
An example of the Hirota derivative is given by 
$$
 D^3_x f(x) \cdot g(x)
   =f(x) (\overleftarrow{\partial_x}-\overrightarrow{\partial_x})^3 g(x).
%\label{2e9}       
$$
The bilinear form with the Hirota derivatives is called as the Hirota form. 
As the special case of Eq.(\ref{2e8}), we consider here the following Hirota type
mKdV equation 
\begin{align}
  (-D_t+D^3_x) f \cdot g&=0    ,
\label{2e10}\\ 
  D^2_x (f \cdot f-g \cdot g)&=0   .
\label{2e11}       
\end{align}
We refer Eqs.(\ref{2e10}) and (\ref{2e11}) as the dynamical equation
and the structure equation, respectively. 
It should be noted that 
in the Hirota type ``KdV'' equation,  it consists of only 
the time-dependent dynamical equation, 
and the  M\"{o}bius group structure is in disguise. 
Thus we consider here the ``mKdV'' equation instead of the KdV equation 
to study the M\"{o}bius group structure for the soliton system.

We work with four different types of forms Eq.(\ref{2e1}), Eq.(\ref{2e2}),
Eq.(\ref{2e6}), Eq.(\ref{2e10}) and  Eq.(\ref{2e11})  for the mKdV
equation.  
Eq.(\ref{2e1}) or Eq.(\ref{2e2}) is the standard mKdV equation.
We use Eq.(\ref{2e6}), and  Eqs.(\ref{2e10}) and (\ref{2e11}) 
to see global  and local M\"{o}bius group structures 
of the mKdV solution, respectively. 

The Schwarzian type equation has nice global  M\"{o}bius group 
(GL(2,$\mathbb{R}$)) symmetry thanks to the Schwarzian derivative.
One can directly show the Schwarzian type equation (\ref{2e6}) is invariant 
under the following 
global  M\"{o}bius transformation, 
\begin{equation}
\varphi(x,t) 
\rightarrow \varphi ' (x,t)=\frac{ \al \varphi(x,t)+\be} {\ga\varphi(x,t)+\de}, 
   \quad (\al, \be, \ga, \de=\text{const.}, \al\de-\be\ga \ne 0),
\label{2e12}
\end{equation}
observing 
\begin{align*}
\big( \varphi ' (x,t)\big)_x  
&=\frac{(\al\de -\be\ga)}{(\ga\varphi(x,t)+\de)^2}\varphi(x,t)_x,
\\
\big( \varphi ' (x,t) \big)_t
&=\frac{(\al\de -\be\ga)}{(\ga\varphi(x,t)+\de)^2}\varphi(x,t)_t,
\\ 
S \big( \varphi ' (x,t), x \big)
&=S(\varphi(x,t), x).
\end{align*}
This global M\"{o}bius group (GL(2,$\mathbb{R}$)) symmetry 
can be decomposed by three symmetries, i.e.\ 
addition formula of $\tanh$, the scale transformation, 
and the translation of $\varphi$: 
\begin{subequations}
\begin{align}
   &a)&  S(\tanh (x+\alpha), x)
&=S \Big( \frac{ \tanh{x}+\tanh{\alpha}}{1+\tanh{x} \tanh{\alpha}}, 
x \Big) 
   =-2, \quad (\alpha=\text{const.}), 
\label{2e16}\\
  &b)& S(\lambda \varphi(x,t), x)&=S(\varphi(x,t), x), \quad 
(\lambda=\text{const.}),
\label{2e17}\\
  &c)& S(\varphi(x,t)+\beta, x)&=S(\varphi(x,t), x), \quad 
(\beta=\text{const.}).
\label{2e18}
\end{align}
\end{subequations}
  
\vspace*{-5mm}
\noindent
In the global M\"{o}bius group (GL(2,$\mathbb{R}$)) symmetry, 
the addition formula of 
the algebraic function ``$\tanh$'' is essential. 
It has been suggested that the addition 
formula of the algebraic function is connected with the 
Lie groups~\cite{Shigemoto1,Shigemoto2}. 

%%%%%%%%%%%%%%%%%%%%%%%%%%%%%%%%%
\subsection{Algebraic construction of the cyclic symmetric
 $N$-soliton solution via local M\"{o}bius group structure}

Through the Schwarzian type mKdV equation, it is clear that there exists the 
global M\"{o}bius group structure in the mKdV equation. 
We next try to construct the cyclic symmetric  $N$-soliton 
solution through the special local ``gauge'' transformation of the full 
M\"{o}bius transformation. 
As we explain in the next section, B\"{a}cklund transformation can be
considered as such a special local gauge transformation.
In the group theoretical approach, there are two ways to construct 
the mKdV cyclic symmetric $N$-soliton solution. 

\vspace*{3mm}

\noindent
{\bf  a) \ Hirota's direct method:} 

\noindent
Though Hirota's method uses the structure 
equation (\ref{2e11}) of the  M\"{o}bius group, 
this method is not algebraic. 
First we solve the structure equation 
(\ref{2e11}) by the {\it perturbation} 
$f=1+\mathcal{O}(\epsilon^2)+\mathcal{O}(\epsilon^4) + \cdots$, \ 
$g=\mathcal{O}(\epsilon)+\mathcal{O}(\epsilon^3) + \cdots$ ,   
for fixed $N$. 
For $N=2$, the addition structure is determined in the form 
\begin{equation}
\tanh{\frac{w}{2}}
=\frac{g}{f}=\frac{\ee^{X_1}+\ee^{X_2}}{1+b_{12}\, \ee^{X_1}
\, \ee^{X_2}},
\label{2e19}
\end{equation}
with $X_i=a_i x+c_i$, $b_{12}=(a_1-a_2)^2/(a_1+a_2)^2$. 
This is the addition structure of the local 
M\"{o}bius group for $N=2$.  
As this addition structure is the local generalization of 
the global M\"{o}bius group structure,
this is quite similar to the addition formula of the algebraic function $\tanh$, 
because if we 
replace $\ee^{X_i} \rightarrow \tanh(\theta_i)$
and $b_{12} \rightarrow 1$, we obtain the addition formula of $\tanh$ itself. 
Then, by finding the 
one-soliton solution of Eq.(\ref{2e10}), we have 
$\tanh(w/2)=g/f=\ee^{a_1 x+a_1^3 t+c_1}$.
As we explain later, it is not necessary to solve the dynamical 
equation (\ref{2e10}) for $N\ge2$. 
By replacing $X_i=a_i x+c_i \rightarrow X_i=a_i x+a_i^3 t+c_i $, 
we obtain the $N=2$ soliton solution. 
The structure equation (\ref{2e11}) is nothing but the 
soliton number preserving B\"{a}cklund transformation itself. 
In the Hirota type mKdV equation, the information of the M\"{o}bius
group structure is build in from the beginning in the form of
Eq.(\ref{2e11}).  
We do not use this method here. See Hirora's textbook~\cite{Hirota4}. 

\vspace*{3mm}

\noindent
  {\bf b) \ B\"{a}cklund transformation method:}

\noindent
This method is the algebraic method. 
The B\"{a}cklund transformation is the special local gauge
transformation of the M\"{o}bius group, which gives new soliton solution
with one increased soliton number. 
This method might be a natural one to understand that
 the mKdV system is the algebraic soliton system. 
By such a B\"{a}cklund transformation, 
we first construct the addition structure such as Eq.(\ref{2e19}).
 In order to obtain the cyclic symmetric $N$-soliton solution 
from the above addition  structure, we simply replace 
$X_i=a_i x+c_i \rightarrow X_i=a_i x+a_i^3 t+c_i $. 
However, as we explain in the next section, 
the obstacle here is that all B\"{a}cklund transformation is not always
commutative. 
In other words, 
the sub-``gauge'' transformation of the non-Abelian M\"{o}bius ``gauge'' 
transformation is not always Abelian.
If the B\"{a}cklund transformation is not commutative, 
the algebraic addition formula does not work to obtain 
the cyclic symmetric  $N$-soliton solution. 

%%%%%%%%%%%%%%%%%%%%%%%%%%%%%   
\subsection{Global M\"{o}bius symmetric mKdV one-soliton 
solutions} 

In order to algebraically construct the cyclic symmetric 
$N$-soliton solution, we first prepare one-soliton solutions.
We here give typical global M\"{o}bius symmetric one-soliton
solutions.

\vspace*{3mm}

\noindent
{\bf a) \ Schwarzian type mKdV solution:}

\noindent
We first clarify an argument of the solution. 
Let us assume that 
$x$ and $t$ come in the combination of $X(x,t)=a x+b t^n+c$. 
Rewriting operators as $\partial_x=a\partial_X$ and 
$\partial_t=bnt^{n-1}\partial_X$, and applying these on functions 
$v$ and $w$ in Eqs.(\ref{2e1}) and (\ref{2e2}), 
it turns out that $n$ should be fixed to be 1, i.e.\ 
$X(x, t)=ax+bt+c$.   

One soliton solution, 
which satisfies the Schwarzian type equation (\ref{2e6}), 
is given by
\begin{equation}
\varphi(x,t)=\varphi(ax-2a^3t+c) =\tanh(ax-2a^3t+c), 
\label{2e20}
\end{equation}
which reproduces the well-known one-soliton solution
\begin{equation}
   v(x,t)=\frac{1}{2}  (\log{\varphi_x})_x =- a \tanh(a x-2a^3 t+c).
\label{2e22} 
\end{equation}
Of course this solution has the global M\"{o}bius group symmetry.
We next examine whether the solution satisfies the Hirota type mKdV equation. 
From Eq.(\ref{2e7}), we have 
\begin{equation}
 \varphi_x=\frac{a}{\cosh^2(ax-2a^3t+c)}=
\ee^{2w}=\frac{(f+g)^2}{(f-g)^2}, 
\label{2e23} 
\end{equation}
which allows $f(x,t)=2\sqrt{a}\, \ee^{X}+\ee^{2X}+1$, 
$g(x,t)=2\sqrt{a}\, \ee^{X}-\ee^{2X}-1$, and $X=a x-2 a^3 t +c$. 
These $f$ and $g$ read 
\begin{eqnarray}
   &&\frac{(-D_t+D^3_x)f \cdot g}{D_x f \cdot g}=3a^2   , 
\label{2e24} \\
   &&\frac{D^2_x (f \cdot f-g \cdot g)}{(f^2-g^2)}=a^2   .
\label{2e25}          
\end{eqnarray}  
While this solution satisfies the mKdV equation (\ref{2e8}), 
it does not satisfy the original Hirota type mKdV equations 
(\ref{2e10}) and (\ref{2e11}), 
but does the generalized Hirota type mKdV equations 
\begin{align}
(-D_t+D^3_x)f \cdot g&=3\lambda^2 D_x f \cdot g, 
\label{2e26} \\
D^2_x (f \cdot f-g \cdot g)&=\lambda^2  (f^2-g^2),
\label{2e27}
\end{align}
with $\lambda=a$. 

\vspace*{3mm}

\noindent
{\bf b) \ Hirota type mKdV solution:}

\noindent
Another well-known one-soliton solution can be derived from 
\begin{equation}
\tanh{\frac{w}{2}}=\frac{g}{f}=\ee^{ ax+a^3 t+c}   .
\label{2e28} 
\end{equation}
We can easily see that this one-soliton solution satisfies 
the Hirota type mKdV equation, 
since 
taking $f=1$, Eqs.(\ref{2e10}) and (\ref{2e11}) are reduced to 
\begin{equation}
-g_t+g_{xxx}=0, 
\quad g_{xx} g -g^2_x=0. 
   \label{2e29}  
\end{equation}
Writing $X=ax+a^3 t +c$, 
Eq.(\ref{2e28}) gives
\begin{equation}
   \ee^{w}=\frac{1+\ee^{X}}{1-\ee^{X}},   
\label{2e30} 
\end{equation}
which reproduces the well-known solution
\begin{equation}
v=w_x=-\frac{a}{\sinh{X}} = -\frac{a}{\sinh(ax+a^3 t +c)}. 
 \label{2e31}                              
\end{equation}

Next we examine whether this solution satisfies 
the Schwarzian type equation or not. 
Having $\varphi_x=\ee^{2w}=(1+\ee^{X})^2/(1-\ee^{X})^2$, 
and integrating this expression by choosing the integration constant properly,  
we could arrive at
\begin{equation}
\varphi=x-\frac{4}{a} \frac{\ee^{X}}{\ee^{X}-1}.   
\label{2e32} 
\end{equation}
Observing 
\begin{equation}
\varphi_x=\frac{\ee^{2X}+2\, \ee^{X}+1}{\ee^{2X}-2\, \ee^{X}+1}, \quad
     \varphi_t=\frac{4 a^2\, \ee^{X}}{\ee^{2X}-2\, \ee^{X}+1},
\label{2e33}                              
\end{equation}
we confirm $\varphi$ given by Eq.(\ref{2e32}) satisfies the Schwarzian type 
mKdV equation (\ref{2e6}). 
We can also check Eq.(\ref{2e32}) reads the same $v(x, t)$ as
Eq.(\ref{2e31})  via 
$v=\frac{1}{2}(\log \varphi_x)_x$. 
One-soliton solution (\ref{2e32}) may have the global M\"{o}bius
symmetry. 

It might be interesting to point out 
that if we choose an integration constant in such a way as 
$$
\displaystyle{\tilde{\varphi}=\frac{X}{a}-\frac{4}{a}
\frac{\ee^{X}}{\ee^{X}-1}}, 
$$ 
we confront the result 
$\tilde{\varphi}_t-\tilde{\varphi}_x S(\tilde{\varphi},x)=a^2$. 
Though $\tilde{\varphi}$ is a solution of 
KdV equation, it is not that of the Schwarzian type equation (\ref{2e6}), 
so that it has no M\"{o}bius group symmetry. 

We have two global  M\"{o}bius symmetric one-soliton 
solutions i.e.\ (\ref{2e20}) and (\ref{2e32}). 
In the next section, we show that 
only the Hirota type mKdV one-soliton solution (\ref{2e32}) is 
connected with the algebraic cyclic symmetric $N$-soliton solution 
through the local M\"{o}bius ``gauge'' transformation, i.e.\ 
B\"{a}cklund transformation. 

%%%%%%%%%%%%%%%%%%%%%%%%%%%%%%%%%%%%%
\section{B\"{a}cklund transformation and construction of 
the cyclic symmetric  $N$-soliton solution} 
\setcounter{equation}{0}

\subsection{B\"{a}cklund transformation of mKdV equation}

The well-known B\"{a}cklund transformation of the mKdV 
equation is given by\cite{Wadati1, Wadati2} 
\begin{align}
w_x'+w_x&=a \sinh(w'-w),
\label{3e1}\\
w_t'+w_t&=-2a^2 w_x-2a w_{xx} \cosh(w'-w)
+(a^3-2a {w_x}^2) \sinh(w'-w).
\label{3e2}\     
\end{align}
This B\"{a}cklund transformation can be considered as 
the special ``gauge'' transformation of the 
M\"{o}bius group~\cite{Crampin}. 
Using the AKNS formalism~\cite{Ablowitz},  
the spacial derivative and its ``gauge'' transformed equation of 
the $2\times 2$ inverse scattering transform are given by 
\begin{align}
\frac{\partial}{\partial x}
  \left(\begin{array}{c}
  \psi_1(x)    \\
  \psi_2(x)
  \end{array}\right)   
&=
  \left(\begin{array}{cc}
  a/2 & v(x) \\
  v(x) & -a/2
  \end{array}\right)
  \left(\begin{array}{c}
  \psi_1(x)    \\
  \psi_2(x)
  \end{array}\right),
\label{3e3}\\
\frac{\partial}{\partial x}
  \left(\lambda(x) \left(\begin{array}{c}
  -\psi_2(x)    \\
  \psi_1(x)
  \end{array}\right)\right)
&= \lambda(x) 
  \left(\begin{array}{cc}
   a/2 & v'(x) \\
   v'(x) & -a/2
  \end{array}\right)
  \left(\begin{array}{c}
   -\psi_2(x)    \\
   \psi_1(x)
  \end{array}\right).
\label{3e4}
\end{align}
Defining $\Gamma=\psi_1/\psi_2$, we have 
$\Gamma_x=a \Gamma+ v (1-\Gamma^2)$  from Eq.(\ref{3e3}).
Using $\Gamma'=\psi'_1/\psi'_2=-\psi_2/\psi_1=-1/\Gamma$ and 
 $\Gamma'_x=a \Gamma'+ v' (1-\Gamma'^2)$, we obtain Eq.(\ref{3e1}) 
 by eliminating $\Gamma$.  

Consistency of $\partial_x \psi_1(x)$ 
and  $\partial_x \psi_2(x)$ in Eqs.(\ref{3e3}) and (\ref{3e4})
gives 
\begin{equation}
(w'_x+w_x)^2=\left((\log{\lambda})_x\right)^2-a^2.
\label{3e5}       
\end{equation}
If we compare the above with the B\"{a}cklund transformation (\ref{3e1}),   
we have 
\begin{equation}
\displaystyle{ (\log{\lambda})_x=a \cosh(w'-w)}. 
\end{equation}
In this way, 
$\lambda(x)$ depends on both $w(x)$ and $w'(x)$. 

We can write this ``gauge'' transformation, which acts on $\psi$, in the form
\begin{equation}
A'=U_x U^{-1}+U A U^{-1}, 
 \label{3e6}
\end{equation}
with 
$$
A=
  \left(\begin{array}{cc}
  a/2 \quad v(x)  \\
  v(x) \ -a/2
  \end{array}\right), \quad  
   A'=
  \left(\begin{array}{cc}
  a/2 \quad v'(x)  \\
  v'(x) \ -a/2
  \end{array}\right), \quad   
  U=\left(\begin{array}{cc}
  0 & -\lambda(x) \\
  \lambda(x) & 0
  \end{array}\right).
\label{3e7}
$$
The B\"{a}cklund transformation (\ref{3e2})
is not necessary to obtain the cyclic symmetric $N$-soliton solution. 
We explain this situation by
using the Hirota type equation of $\tanh(w/2)=g/f$.
The B\"{a}cklund transformation of the Hirota type, which corresponds 
to  Eq.(\ref{3e1}), is given by~\cite{Hirota3}
\begin{equation}
D_x (f \pm g) \cdot (f' \mp g')
      =\frac{a}{2} (f \mp g) (f' \pm g'),
\label{3e8} 
\end{equation}
which gives
\begin{equation}
D^2_x (f \pm g) \cdot (f'\pm g')
      =\frac{a^2}{4} (f\pm g) (f'\pm g').
\label{3e9}               
\end{equation}
If we write the B\"{a}cklund transformation in the Hirota form,
the B\"{a}cklund transformation (\ref{3e9}) and the generalized structure 
equation (\ref{2e27}) becomes strongly related in that form.
Using Eqs.(\ref{2e11}) and (\ref{3e8}), we can show 
$ D^2_x (f' \cdot f'-g' \cdot g')=0$~\cite{Hirota3}.  
This means that if we construct $f'$ and
$g'$ from $f$ and $g$, which satisfy  Eq.(\ref{3e8}), $f'$ and $g'$ 
automatically satisfy the Hirota type dynamical equation 
$(-D_t+D^3_x)f' \cdot g'=0$ by 
using the primed bilinear Hirota form (\ref{2e8})   
\begin{equation}
\frac{(-D_t+D^3_x)f' \cdot g'}{D_x f' \cdot g'}
   =3\cdot\frac{{D^2}_x (f' \cdot f'-g' \cdot g')}{({f'}^2-{g'}^2)}.
\label{3e10}       
\end{equation}
We use the Hirota type dynamical equation (\ref{2e10}) only 
in the case of solving the one-soliton solution. 
 
If $\tanh(w/2)=g/f$ is the mKdV solution, 
$\tanh(-w/2)=-g/f$ is also a solution.
Then the soliton number preserving B\"{a}cklund transformation 
is given by $a=0$, $f'=f$, $g'=-g$ in Eqs.(\ref{3e8}) and (\ref{3e9}). 
Eq.(\ref{3e9}) in this case is given as 
\begin{equation}
  D^2_x (f +g) \cdot (f -g)=D^2_x (f \cdot f -g \cdot g)=0.
\label{3e11}               
\end{equation}
This is nothing but the Hirota type structure equation Eq.(\ref{2e11}). 
   
The problem of the above B\"{a}cklund transformation is 
that this B\"{a}cklund transformation is not commutative. 
We show this by reductio ad absurdum.
Assuming the commutativity 
$w_{12}=w_{21}$, we have 
\begin{subequations} 
\begin{align}
w_{1, x}+w_{0, x}&=a_1 \sinh(w_{1}-w_{0}),
\label{3e12}\\
w_{2, x}+w_{0, x}&=a_2 \sinh(w_{2}-w_{0}),
\label{3e13}\\
w_{12, x}+w_{1, x}&=a_2 \sinh(w_{12}-w_{1}),
\label{3e14}\\  
w_{12, x}+w_{2, x}&=a_1 \sinh(w_{12}-w_{2}).
\label{3e15}   
\end{align} 
\end{subequations}

\vspace*{-5mm}
\noindent
Manipulating 
Eq.(\ref{3e12})$-$Eq.(\ref{3e13})$-$Eq.(\ref{3e14})$+$Eq.(\ref{3e15}), 
the derivative terms are canceled out, 
so that we have an algebraic relation
\begin{equation}
\tanh\Big(\frac{w_{12}-w_{0}}{2}\Big)=-\frac{a_1+a_2}{a_1-a_2} 
   \tanh\Big(\frac{w_{1}-w_{2}}{2}\Big).  
\label{3e16}
\end{equation}
We can check that Eq.(\ref{3e16}) satisfies 
Eqs.(\ref{3e12})-(\ref{3e15}), which means that it is 
commutative in this level.
We could explain the non-commutativity of this B\"{a}cklund transformation 
by constructing the three-soliton solution.
Putting $w_0=0$, we have 
\begin{align}
\tanh\Big(\frac{w_{12}}{2}\Big)&=
-a_{12} \tanh\Big(\frac{w_{1}-w_{2}}{2}\Big)=
 -a_{12} \frac{\tanh(w_{1}/2)-\tanh(w_{2}/2)}
{1-\tanh(w_{1}/2) \tanh(w_{2}/2)},
\label{3e17}\\
\tanh\Big(\frac{w_{13}}{2}\Big)&=
-a_{13} \tanh\Big(\frac{w_{1}-w_{3}}{2}\Big)=
 -a_{13} \frac{\tanh(w_{1}/2)-\tanh(w_{3}/2)}
{1-\tanh(w_{1}/2) \tanh(w_{3}/2)},
\label{3e18}
\end{align}
with $a_{ij}=(a_i+a_j)/(a_i-a_j)$. 
Next, let us construct a three-soliton solution. 
Assuming the part of the commutativity
$w_{123}=w_{132}$, we have
\begin{subequations}
\begin{align}
w_{12, x}+w_{1, x}&=a_2 \sinh(w_{12}-w_{1}),
\label{3e19}\\ 
w_{13, x}+w_{1, x}&=a_3 \sinh(w_{13}-w_{1}),   
\label{3e20} \\
w_{123, x}+w_{12, x}&=a_3 \sinh(w_{123}-w_{12}),
\label{3e21}\\
w_{123, x}+w_{13, x}&=a_2 \sinh(w_{123}-w_{13}).
\label{3e22}
\end{align}
\end{subequations}

\vspace*{-5mm}
\noindent 
Making Eq.(\ref{3e19})$-$Eq.(\ref{3e20})$-$Eq.(\ref{3e21})$+$
Eq.(\ref{3e22}), an algebraic relation shows up 
\begin{equation}
\tanh\Big(\frac{w_{123}-w_{1}}{2}\big)=-\frac{a_2+a_3}{a_2-a_3} 
   \tanh\Big(\frac{w_{12}-w_{13}}{2}\Big)=A.
\label{3e23}
\end{equation}
We have $\tanh(w_{123}/2)
=(A+\tanh(w_{1}/2) )/(1+A\tanh(w_{1}/2))$.
Our question is, by using the addition formula Eqs.(\ref{3e17}), 
(\ref{3e18}), and (\ref{3e23}),  
whether $\tanh(w^{(123)}/2)$ becomes the cyclic 
symmetric expression or not. 
This can be the check of the full commutativity in this case. 
The answer is in the negative, 
since we have the following expression without cyclic
symmetry for $ t_1=\tanh(w_{1}/2)$,
$t_2=\tanh(w_{2}/2)$, and $t_3=\tanh(w_{3}/2)$:   
\begin{equation}
\tanh\frac{w_{123}}{2}=
\frac{\tanh(w_{1}/2)
\big(1-\tanh(w_{12}/2) \tanh(w_{13}/2)\big)
-a_{23}\big(\tanh(w_{12}/2)-\tanh(w_{13}/2)\big)}
{1-\tanh(w_{12}/2)  \tanh(w_{13}/2) 
-a_{23} \tanh(w_{1}/2)
\big(\tanh(w_{12}/2)-\tanh(w_{13}/2)\big)}, 
\label{3e24}
\end{equation}
with Eqs.(\ref{3e17}) and (\ref{3e18}). 
The situation differs for the well-known B\"{a}cklund transformation of 
KdV equation and this B\"{a}cklund transformation becomes commutative. 
We see that in the next section. 

%%%%%%%%%%%%%%%%%%%%%%%%%%%%%%%
\subsection{B\"{a}cklund transformation of KdV equation}

Since we would like to construct 
the mKdV cyclic symmetric $N$-soliton solution through the KdV equation, 
we here review the KdV equation. 

Using the variable $u=z_x$, the standard KdV 
equation is given by
\begin{align}
u_t-u_{xxx}+6 u u_x&=0,
\label{3e25}\\
z_t-z_{xxx}+3 {z_x}^2&=0, 
\label{3e26} 
\end{align}  
where we put an integration constant to be zero.
We use the following Hirota type KdV equation~\cite{Hirota1} 
\begin{align}
&(-D_t D_x+D^4_x) \tau \cdot \tau=0,
\label{3e27}\\ 
&u=-2 (\log{\tau})_{xx}=-2 \left( \frac{\tau_x}{\tau} \right)_x. 
\label{3e28} 
\end{align}
The well-known  B\"{a}cklund transformation of the KdV 
equation~\cite{Wahlquist} is given by 
\begin{align}
z'_x+z_x&=-\frac{a^2}{2}+\frac{(z'-z)^2}{2},
\label{3e29} \\
z'_t+z_t&=-2({z'_x}^2+z'_x z_x+ z^2_x)+(z'-z)(z'_{xx}-z_{xx}).
\label{3e30} 
\end{align}  
It should be mentioning that 
the B\"{a}cklund transformations above are invariant under 
the constant shift $z\rightarrow z+c_0$, $z' \rightarrow z'+c_0$, 
($c_0={\rm const.}$), which we use later. 
This B\"{a}cklund transformation is the special ``gauge'' transformation of the 
M\"{o}bius group~\cite{Crampin}.
Using the AKNS formalism~\cite{Ablowitz},  
the spacial derivative equation and its ``gauge'' transformed 
equation of the $2 \times 2$ inverse scattering transform are given by 
\begin{align}
\frac{\partial}{\partial x}
  \left(\begin{array}{c}
  \psi_1(x)    \\
  \psi_2(x)
  \end{array}\right)  &=
  \left(\begin{array}{cc}
  a/2 & -u(x) \\
  -1 & -a/2
  \end{array}\right)
  \left(\begin{array}{c}
  \psi_1(x)    \\
  \psi_2(x)
  \end{array}\right),
\label{3e31}\\
\frac{\partial}{\partial x}
  \left(\lambda(x) \left(\begin{array}{c}
  \psi_1(x)+a \psi_2(x)    \\
  -\psi_2(x)
  \end{array}\right)\right)&=\lambda(x) 
  \left(\begin{array}{cc}
   a/2 & -u'(x) \\
   -1 & -a/2
  \end{array}\right)
  \left(\begin{array}{c}
   \psi_1(x)+a \psi_2(x)    \\
    -\psi_2(x)
  \end{array}\right).
\label{3e32}
\end{align}
As we did for the case of mKdV equations, 
defining $\Gamma=\psi_1/\psi_2$ then we have 
$\Gamma_x=-u+a \Gamma +\Gamma^2$ from Eq.(\ref{3e31}).
By using $\Gamma'=\psi'_1/\psi'_2=-(\psi_1+a \psi_2)/\psi_2=-\Gamma-a$ and 
 $\Gamma'_x=-u'+a \Gamma' +\Gamma'^2$,  
we obtain Eq.(\ref{3e29}) by eliminating $\Gamma$. 

Consistency of $\partial_x \psi_1(x)$ and  $\partial_x \psi_2(x)$ in 
Eqs.(\ref{3e31}) and (\ref{3e32})
gives 
\begin{equation}
z'_x+z_x=-\frac{a^2}{2}+\frac{(\partial_x \lambda/\lambda)^2}{2}.
\label{3e33}       
\end{equation}
If we compare the above with the structure KdV 
B\"{a}cklund transformation (\ref{3e29}), 
we have 
\begin{equation}
 \displaystyle{ (\log{\lambda})_x=- (z-z')}.
\end{equation}
 We can write this ``gauge'' transformation, which acts on $\psi(x)$, in the form
\begin{equation}
A'=U_x U^{-1}+U A U^{-1}, 
 \label{3e34}
\end{equation}
with 
$$   A=
  \left(\begin{array}{cc}
  a/2 \quad -u(x)  \\
  -1 \ -a/2
  \end{array}\right), \quad  
   A'=
  \left(\begin{array}{cc}
  a/2 \quad -u'(x)  \\
  -1 \ -a/2
  \end{array}\right), \quad   
  U=\left(\begin{array}{cc}
  \lambda(x)  & a \lambda(x) \\
  0  & -\lambda(x) 
  \end{array}\right).
$$

The Hirota type B\"{a}cklund transformation,
which corresponds to  Eq.(\ref{3e29}), is given by~\cite{Hirota3}
\begin{equation}
D^2_x  \tau' \cdot \tau =\frac{a^2}{4} \tau' \tau.
\label{3e36}
\end{equation}

%%%%%%%%%%%%%%%
\subsection{Connection between KdV and mKdV equations}

The connection between the mKdV equation and the KdV equation is given 
by the Miura transformation in the form $u=\pm v_x +v^2$, 
\begin{equation}
u_t-u_{xxx}+6 u u_x =\pm (\partial_x \pm 2 v)( v_t-v_{xxx}+6 v^2 v_x).
\label{3e37}
\end{equation}
First, we decompose the $\tau$ function of KdV equation into the even 
part and the odd part. 
As an example, we consider the two-soliton solution of KdV equation
\begin{equation}
   \tau(x,t)=1+\ee^{X_1}+\ee^{X_2}+b_{12}\, \ee^{X_1}\ee^{X_2}  .
\label{3e38} 
\end{equation}
We define the even and odd part of $\tau$ as behavior under 
$\ee^{X_i} \rightarrow -\ee^{X_i} $.  
We then decompose the even part by
$f_1=1+b_{12}\, \ee^{X_1}\ee^{X_2}$ 
and the odd part by $g_1=\ee^{X_1}+\ee^{X_2}$, which gives
 $\tau=f_1+g_1$. 
In this decomposition, $u$ is expressed by 
\begin{equation}
u=-2(\log{\tau})_{xx}=-2 (\log(f_1+g_1))_{xx}=-2\partial_x 
  \left(   \frac{ f_{1, x}+g_{1, x}}{f_1+g_1}\right).
\label{3e39} 
\end{equation} 
In terms of mKdV equation, using $\ee^{w}=(f+g)/(f-g)$ of Eq.(\ref{2e7}), $v$ 
is expressed by 
\begin{equation}
v=w_x=\frac{2(f g_x- f_x g)}{f^2-g^2}.
\label{3e40} 
\end{equation}
Surprisingly, if we identify $f_1=f$ and $g_1=g$, 
we can show that the Miura transformation $u=-v_x+v^2$
gives the Hirota type structure equation (\ref{2e11}) itself
by using the relation $0=(f^2-g^2)(-u-v_x+v^2)=D^2_x(f \cdot f - g \cdot g)$. 
Therefore, as Eqs.(\ref{2e10}) and (\ref{2e11}) are satisfied,
the solution of mKdV equation $v=w_x$ with $\tanh(w/2)=g/f$
 corresponds to the solution of KdV equation $u=-2(\log{\tau})_{xx}$ 
with $\tau=f+g$ through the Miura transformation in the following form
 \begin{eqnarray}
  \text{mKdV equation}  &&\xleftrightarrow{\ \text{Miura tr.}\, } 
\quad \text{KdV equation} 
  \nonumber\\
  v=w_x,\quad  \tanh{\frac{w}{2}}=\frac{g}{f} && \hspace{-2mm} 
  \xleftrightarrow{\ \text{$u=-v_x+v^2$}\, }  
\quad u=-2(\log{\tau})_{xx},\quad \tau=f+g.    
\label{3e41}                               
\end{eqnarray}
 That is, if we know the KdV solution, we can  
 obtain the mKdV solution with the same $f$ and $g$ and vice versa.

 For the KdV equation, if $\tau=f+g$ is the solution with $f$ being 
 the even function and $g$ being the odd 
 function under $\ee^{X_i} \rightarrow -\ee^{X_i} $, $\tau'=f-g$ 
is also the solution. 
Then the soliton number preserving B\"{a}cklund 
 transformation is given by putting $a=0$, $\tau=f+g$,  $\tau'=f-g$ in 
 Eq.(\ref{3e36}), which gives 
 \begin{equation}
D^2_x (f+g) \cdot (f-g)=D^2_x (f\cdot f-g\cdot g) =0.
\label{3e42}
\end{equation}
We again obtain the Hirota type structure equation.
The Hirota type equation is not unique for the given standard 
type soliton equation. 
For the standard type mKdV equation (\ref{2e1}), 
 there are various different Hirota type equation by assuming $v=g/f$ or 
 $w=\log(g/f)$ or $\tanh(w/2)=g/f$~\cite{Hirota2,Hirota4}. 
The Miura transformation 
 $u=-v_x+v^2$ from $v$ to $u$ 
 is easily performed by the differentiation, but the inverse Miura transformation
 from $u$ to $v$ is in general quite difficult because we must solve the 
 non-linear differential equation.  
In order to algebraically find the cyclic symmetric $N$-soliton solution of 
the mKdV equation, this Miura transformation must also be algebraic.
In order to do so, we must use the special Hirota type equation 
coming from $\tanh(w/2)=g/f$.
In that case, the Miura transformation connects mKdV 
equation and KdV equation  just only by the correspondence 
of the same $f$ and $g$ with $\tanh(w/2)=g/f$ in mKdV 
and $\tau=f+g$ in KdV, which means that the Miura transformation 
and the  inverse Miura transformation is algebraic. 

Furthermore, if we use the special Hirota type mKdV equation coming from
 $\tanh(w/2)=g/f$, the Miura transformation becomes equivalent to the 
 soliton number preserving KdV B\"{a}cklund
 transformation (\ref{3e42}). 
Then such mKdV Hirota type structure equation has Abelian group structure. 
This is because $i)$ Hirota type structure 
 equation is the same as the soliton number preserving KdV B\"{a}cklund
 transformation, $ii)$ KdV B\"{a}cklund transformation has Abelian group 
 structure. 
The Abelian group structure of this Hirota type structure 
equation is the reason why the Hirota's direct method works
well for such a special Hirota type equation, where it is not necessary to 
connect the mKdV equation and the KdV  equation.  

%%%%%%%%%%%%%%%%%%%%%%%%
\subsection{Three soliton solution - 
The demonstration of construction of the mKdV cyclic symmetric 
$N$-soliton solutions}

We demonstrate to construct three-soliton solution, 
which is the nontrivial case to demonstrate 
the commutativity of the B\"{a}cklund transformation. 
We start from 3 one-soliton 
solutions of the Hirota type equation, 
\begin{equation}
\tanh{\frac{w_i}{2}}=\frac{g_i}{f_i}, \quad f_i=1, \quad g_i =\ee^{X_i}  ,  
\label{3e43}
\end{equation}
with
$$
X_i=a_i x+a^3_it +c_i, \ (i=1,2,3),  \quad 
v_i=-\frac{a_i}{\sinh X_i}.
$$
From the Miura transformation, we have 
\begin{align}
u_i&=z_{i, x}=-v_{i, x}+v^2_{i} = - \frac{a^2_i}{2} \frac{1}{\cosh^2(X_{i}/2)},
\label{3e46} \\
z_i&=-a_i \tanh \frac{X_{i}}{2} =a_i \frac{(1-\ee^{X_i})}{(1+\ee^{X_i})}.
\label{3e47}    
 \end{align}
Let us first construct two-soliton solution 
by the B\"{a}cklund transformation (\ref{3e29}).
Assuming the commutativity, $z_{12}=z_{21}$, we have
\begin{subequations} 
\begin{align}
 z_{1, x}+z_{0, x}&=-\dfrac{a^2_1}{2}+\dfrac{(z_1-z_0)^2}{2},
\label{3e48}\\
z_{2, x}+z_{0, x}&=-\dfrac{a^2_2}{2}+\dfrac{(z_2-z_0)^2}{2},
\label{3e49}\\
z_{12, x}+z_{1, x}&=-\dfrac{a^2_2}{2}+\dfrac{(z_{12}-z_1)^2}{2},
\label{3e50}\\
z_{12, x}+z_{2, x}&=-\dfrac{a^2_1}{2}+\dfrac{(z_{12}-z_2)^2}{2}.
\label{3e51}
\end{align}  
\end{subequations}

\vspace*{-5mm}
\noindent 
As we did before, 
making Eq.(\ref{3e48})$-$Eq.(\ref{3e49})$-$Eq.(\ref{3e50})$+$Eq.(\ref{3e51}),
the derivative terms are canceled out and we have
$z_{12}-z_0=(a^2_1-a^2_2)/(z_1-z_2)$. 
Putting $z_0=0$,  $z_{12}$, $z_{13}$ are given by 
\begin{equation}
z_{12}=\frac{a^2_1-a^2_2}{z_1-z_2}, \quad 
z_{13}=\frac{a^2_1-a^2_3}{z_1-z_3}. 
\label{3e52}   
\end{equation}
We can check that Eq.(\ref{3e52}) satisfies 
Eqs.(\ref{3e48})-(\ref{3e51}), which means that it is 
commutative in this level. 

Next let us construct the three-soliton solution. 
Assuming the part of the commutativity, $z_{123}=z_{132}$,  we have 
%, 
\begin{subequations}
\begin{align}
z_{12, x}+z_{1, x}&=-\frac{a^2_2}{2}+\frac{(z_{12}-z_1)^2}{2},
\label{3e53}\\
z_{13, x}+z_{1, x}&=-\frac{a^2_3}{2}+\frac{(z_{13}-z_1)^2}{2},
\label{3e54}\\
z_{123, x}+z_{12, x}&=-\frac{a^2_3}{2}+\frac{(z_{123}-z_{12})^2}{2},
\label{3e55}\\
z_{123, x}+z_{13, x}&=-\frac{a^2_2}{2}+\frac{(z_{123}-z_{13})^2}{2}.
\label{3e56}
\end{align} 
\end{subequations}

\vspace*{-5mm}
\noindent
Making Eq.(\ref{3e53})$-$Eq.(\ref{3e54})$-$
Eq.(\ref{3e55})$+$Eq.(\ref{3e56}), we obtain 
\begin{align}
z_{123}&=z_1+\frac{a^2_2-a^2_3}{z_{12}-z_{13}}  \nonumber\\
   &=-\frac{(a^2_1-a^2_2)z_1 z_2+(a^2_2-a^2_3)z_2 z_3+(a^2_3-a^2_1)z_3 z_1}
      {(a^2_1-a^2_2)z_3+(a^2_2-a^2_3)z_1+(a^2_3-a^2_1)z_2}. 
\label{3e57}
\end{align}
We can check that Eq.(\ref{3e57}) really satisfies  
Eqs.(\ref{3e53})-(\ref{3e56}), which means that it is partly commutative in this level.
In the above, we use only $z_{12}$, $z_{13}$ and do not use $z_{23}$. 
However, as the result,  
$z_{123}$ is cyclic symmetric in $z_1, z_2, z_3$, which is 
the check of the full commutativity of this B\"{a}cklund transformation.
See also Bianchi and Eisenhart\cite{Bianchi, Eisenhart}. 

Using Eq.(\ref{3e47}) and redefine the constants $c_i \ (i=1,2,3)$ in
such a way as 
$\hat{X}_i=a_i x+a^3_i t +\hat{c}_i$ with 
\begin{equation}
\ee^{\hat{X}_1}
=\frac{(a_1+a_2)(a_1+a_3)}{(a_1-a_2)(a_1-a_3)}\, \ee^{X_1}, \ \ 
  \ee^{\hat{X}_2}=\frac{(a_2+a_1)(a_2+a_3)}{(a_2-a_1)(a_2-a_3)}\, \ee^{X_2}, 
\ \
\ee^{\hat{X}_3}
=\frac{(a_3+a_1)(a_3+a_2)}{(a_3-a_1)(a_3-a_2)}\, \ee^{X_3}, 
\label{3e58}
\end{equation}
this gives
\begin{equation}
z_{123}=(a_1+a_2+a_3)-2\frac{\tau_x}{\tau}, 
\label{3e59}
\end{equation}
where 
\begin{align}
\tau&=1+\ee^{\hat{X}_1}+\ee^{\hat{X}_2}+\ee^{\hat{X}_3}
+b_{12}\, \ee^{\hat{X}_1+\hat{X}_2}
+b_{13}\, \ee^{\hat{X}_1+\hat{X}_3}
+b_{23}\, \ee^{\hat{X}_2+\hat{X}_3} \nonumber\\
   &\hspace*{4mm}
+b_{12} b_{13} b_{23}\, \ee^{\hat{X}_1+\hat{X}_2+\hat{X}_3}, 
\label{3e60}
\end{align}
with
$$
b_{ij}=\frac{(a_i-a_j)^2}{(a_i+a_j)^2}. 
\label{3e61}
$$
The constructed $z_{123}$ 
differs from $-2 \tau_x/\tau$  by the constant factor $(a_1+a_2+a_3)$, 
but this constant factor does not contribute to
$u=z_{123, x}= -2\left( \tau_x/\tau \right)_x $. 
This is equivalent to the constant shift of 
$z_{123} \rightarrow   z_{123}+(a_1+a_2+a_3)$. 
Then we have even and odd part of $\tau$ in the form
\begin{align}
f&=1+b_{12}\, \ee^{\hat{X}_1+\hat{X}_2}
+b_{13}\, \ee^{\hat{X}_1+\hat{X}_3}
   +b_{23}\, \ee^{\hat{X}_2+\hat{X}_3},
\label{3e62}\\
g&=\ee^{\hat{X}_1}+\ee^{\hat{X}_2}+\ee^{\hat{X}_3}
     +b_{12} b_{13} b_{23}\, \ee^{\hat{X}_1+\hat{X}_2+\hat{X}_3}. 
\label{3e63}
\end{align} 
This gives the mKdV three-soliton solution
\begin{equation}
\tanh{\frac{w}{2}}
=\frac{g}{f}=\frac{\ee^{\hat{X}_1}+\ee^{\hat{X}_2}+\ee^{\hat{X}_3}
     +b_{12} b_{13} b_{23}\, \ee^{\hat{X}_1+\hat{X}_2+\hat{X}_3} }
     {1+b_{12}\, \ee^{\hat{X}_1+\hat{X}_2}+b_{13}\, 
\ee^{\hat{X}_1+\hat{X}_3}
     +b_{23}\, \ee^{\hat{X}_2+\hat{X}_3}}.
\label{3e64}
\end{equation}
This addition structure is quite similar to the addition formula of
$\tanh$, 
since we obtain the addition formula of $\tanh$ by the replacement 
$\ee^{\hat{X}_i} \rightarrow \tanh(\theta_i)$ and $b_{ij} \rightarrow 1$. 

The general cyclic symmetric $N$-soliton solution obtained by the Hirota's direct 
method which is non-algebraic method is given in Hirota's paper~\cite{Hirota2}. 

%%%%%%%%%%%%%%%%%%%%%%%%%%%%%%%%%%%%%%%%%%%%%%%%
\section{Summary and discussions} 
\setcounter{equation}{0}

We have the dogma that it is quite surprising that the soliton equation 
has infinitely many exact solutions 
($N$-soliton solution with $N \rightarrow \infty$) 
despite of its non-linearity. 
There must exist some nice structures  
and they must be some local Lie group structures behind the non-linear
soliton equation.  
According to our dogma,  we study to construct the mKdV cyclic symmetric
 $N$-soliton solution and we elucidate the mechanism why we can algebraically  
 construct the cyclic symetric $N$-soliton solution 
from the group theoretical point of view.  

First of all, the Schwarzian type mKdV equation has the global 
M\"{o}bius group (GL(2,$\mathbb{R}$)) symmetry. 
It is natural to expect that there might be the local
M\"{o}bius group (GL(2,$\mathbb{R}$)) symmetry. 
We then try to construct the cyclic symmetric  
$N$-soliton solution algebraically from one-soliton 
solutions through the B\"{a}cklund transformation.
The B\"{a}cklund transformation is considered 
as the local M\"{o}bius transformation,
but the M\"{o}bius group is generally non-Abelian, so that only 
some special B\"{a}cklund transformation is commutative for the 
addition of the transformations. 
While the well-known mKdV B\"{a}cklund 
transformation is not commutative, the well-known 
KdV B\"{a}cklund transformation is commutative. 
Then, in order to construct the mKdV cyclic symmetric 
$N$-soliton solution by the algebraic addition formula 
of the B\"{a}cklund transformation, 
we first transform the mKdV equation to the KdV equation 
by the Miura transformation, 
and use the KdV B\"{a}cklund transformation 
to obtain the KdV cyclic symmetric $N$-soliton solution from one-soliton 
solutions and finally come back to the mKdV cyclic symmetric $N$-soliton solution 
by the inverse Miura transformation.
For $N=2$ case, we use the following scheme:  
\begin{eqnarray}
  &&\tanh{\frac{w_i}{2}}=\ee^{X_i} \ \xrightarrow{\text{Miura tr.}} \ 
  z_i=a_i\frac{(1-\ee^{X_i})}{(1+\ee^{X_i})}\ 
\xrightarrow{\text{B\"{a}cklund tr.}} \ 
  z_{12}=\frac{a^2_1-a^2_2}{z_1-z_2}=-2\frac{\tau_x}{\tau} +a_1+a_2 \nonumber\\
  &&\hspace{15mm} \Uparrow \hspace{30mm}  
  \Uparrow  \hspace{50mm} \Uparrow \nonumber\\
  &&\hspace{-1mm}\text{mKdV one-soliton sol.} \hspace{5mm} 
\text{KdV one-soliton sol.} 
  \hspace{15mm}
   \text{KdV two-soliton sol.}
  \nonumber\\
   && \nonumber\\
    &&\hspace{-10mm} \rightarrow 
\tau=1+\ee^{\hat{X}_1}+\ee^{\hat{X}_2}
+b_{12}\, \ee^{\hat{X}_1}\, \ee^{\hat{X}_2}=f+g
     \ \xrightarrow{\text{inverse Miura tr.}} \ 
      \tanh{\frac{w_{12}}{2}}=\frac{g}{f}
=\frac{\ee^{\hat{X}_1}+\ee^{\hat{X}_2}}
{1+b_{12}\, \ee^{\hat{X}_1}\, \ee^{\hat{X}_2}} 
   \nonumber\\
  &&\hspace{-1mm} \Uparrow \hspace{106mm} \Uparrow
  \nonumber\\
  &&\hspace{-5mm} \text{KdV two-soliton $\tau$\ funct. } 
\hspace{60mm} \text{mKdV two-soliton sol.}
  \nonumber                               
\end{eqnarray} 

However, if we algebraically construct the cyclic symmetric
 $N$-soliton solution, the above B\"{a}cklund transformation 
and the inverse B\"{a}cklund transformation must be algebraic.
If we use the special type of Hirota equation which comes 
from $\tanh(w/2)=g/f$, as it is quite surprising, this Miura transformation
 becomes equivalent to the structure equation of this Hirota 
 type mKdV equation (\ref{2e11}) and also becomes 
equivalent to the soliton number preserving 
KdV B\"{a}cklund transformation. 
Therefore, the local M\"{o}bius group structure is already build in as 
the structure equation of this Hirota type mKdV equation
and this structure equation has the Abelian group structure.
This is the reason why the Hirota's direct method works well, where 
it is not necessary to connection the mKdV equation and the 
KdV  equation.

In order that the whole construction of the cyclic symmetric
 $N$-soliton solution becomes algebraic, the one-soliton 
solution must be some special one-soliton solution, that is, it must be 
the one-soliton solution of the special Hirota type equation which 
comes from $\tanh(w/2)=g/f$. 
Regarding the addition structure of $\tanh(w/2)=g/f$, we have 
Eq.(\ref{2e19}) for $N=2$ and Eq.(\ref{3e64}) for $N=3$, which are quite 
similar to the addition formula of $\tanh$, which reflect the addition formula 
of the global  M\"{o}bius group symmetry of the Schwarz type equation,
where the addition formula of $\tanh$ is essential.
The global addition structure is given by the global transformation from 
one-soliton solution to another one-soliton solution in the following form: \\

i) \, $w \rightarrow w'=w+c$, which gives  another 
one-soliton solution  
$$\displaystyle{\tanh(w'/2)=\frac{\tanh(w/2)
+\tanh(c/2)}{1+\tanh(w/2)\tanh(c/2)}}\,, $$%\\

ii) \, $c_1 \rightarrow c_1 +\log \lambda$, which gives another 
one-soliton solution 
\[\tanh(w'/2)=\lambda \ee^{X_1},\] 
which is the 
scale transformation. %\\
Combining i) and ii), we have the global  M\"{o}bius group symmetry. 

We can apply our method to the cyclic symmetric
 $N$-soliton solution of the 
$\sinh$-Gordon equation $\theta_{xt}=\sinh\theta$. The  
B\"{a}cklund transformation of the $\sinh$-Gordon equation 
 $\theta'_x/2+\theta_x/2=a\sinh(\theta'/2-\theta/2)$
has the same form as Eq.(\ref{3e1}) in mKdV equation. 
Then, by putting $\tanh(\theta/4)=g/f$ and $X_i=a_i x+t/a_i+ c_i$,
we obtain the Hirota's result \cite{Hirota5} by using our method.

\vspace{10mm}

%%%%%%%%%%%%%%%%%%%%%%%%%%%%%%%%%%%%%%%%%%%%%%


\begin{thebibliography}{99}
\bibitem{Gardner}
C.S. Gardner, J.M. Greene, M.D. Kruskal, and R.M. Miura, 
Phys. Rev. Lett. {\bf 19}, 1095 (1967).
\bibitem{Lax} 
P.D. Lax, Commun, Pure and Appl. Math. {\bf 21}, 467 (1968).
\bibitem{Zakhrov}
V.E. Zakharov and A.B. Shabat, Sov. Phys. JETP {\bf 34}, (1972) 62.
\bibitem{Ablowitz}
M.J. Ablowitz, D.J. Kaup, A.C. Newell, and H. Segur, 
Phys. Rev. Lett. {\bf 31}, 125 (1973).  
\bibitem{Sasaki}
R. Sasaki, Nucl. Phys. {\bf B154}, 343 (1979).  
\bibitem{Wahlquist}
H.D. Wahlquist and F.B. Estabrook, Phys. Rev. Lett. {\bf 31}, 1386 (1973). 
\bibitem{Wadati1} 
M. Wadati, J. Phys. Soc. Jpn. {\bf 36}, 1498 (1974). 
\bibitem{Wadati2}
K. Konno and M. Wadati, Prog. Theor. Phys. {\bf 53}, 1652 (1975). 
\bibitem{Hirota3}
R. Hirota, Prog. Theor. Phys. {\bf 52}, 1498 (1974). 
\bibitem{Hirota1}
R. Hirota, Phys. Rev. Lett. {\bf 27}, 1192 (1971). 
\bibitem{Hirota2}
R. Hirota, J. Phys. Soc. Jpn. {\bf 33}, 1456 (1972).
\bibitem{Hirota4}
R. Hirota, {\it Direct Method in Soliton Theory} (Cambridge 
University Press, Cambridge, 2004). 
\bibitem{Sato}
M. Sato, RIMS Kokyuroku (Kyoto University) {\bf 439}, 30 (1981).
\bibitem{Lepowsky}
J. Lepowsky and R.L. Wilson, Comm. Math. Phys. {\bf 62}, 43 (1978).
\bibitem{Date}
E. Date, M. Kashiwara, and T. Miwa, Proc. Japan Acad. {\bf 57A}, 387 (1981).
\bibitem{Weiss}
J. Weiss, J. Math. Phys. {\bf 24}, 1405 (1983). 
\bibitem{Shigemoto1}
K. Shigemoto, Tezukayama Academic Review {\bf 17}, 15 (2011).
\bibitem{Shigemoto2}
K. Shigemoto, Tezukayama Academic Review {\bf 19}, 1 (2013).
\bibitem{Crampin}
M. Crampin, Phys. Lett. {\bf A66}, 170 (1978).
\bibitem{Bianchi}
L. Bianchi, {\it Vorlesungen \"{u}ber Differenzialgeometrie} (Teubner, 1899), p.418. 
\bibitem{Eisenhart}
L.P. Eisenhart, {\it A Treatice on the Differential Geometry 
of Curves and Surfaces} (Dover, New York, 1960), p.286. 
\bibitem{Hirota5}
R. Hirota, J. Phys. Soc. Jpn. {\bf 33}, 1459 (1972).
\end{thebibliography}
\end{document}